\documentclass[aps,pra,superscriptaddress,showpacs,twocolumn]{revtex4-1}
\usepackage{amsmath,amsfonts,amssymb,amstext,amscd,amsthm,dsfont}
\usepackage{mathrsfs}
\usepackage{bbold}
\usepackage{mathtools,color}
\usepackage[mathscr]{eucal}
\usepackage{amssymb}
\usepackage{graphics}
\usepackage{hyperref}
\usepackage{blindtext}
\usepackage[all,cmtip]{xy}
\newcommand\ket[1]{\left|#1\right>}
\newcommand\bra[1]{\left<#1\right|}
\newcommand\braket[2]{\left<#1\right|\left.\!\!#2\right>}
\newcommand{\ketbra}[2]{\ket{#1}\!\bra{#2}}
\newcommand\projj[2]{\left|#1\right>\!\left<#2\right|}
\newcommand\proj[1]{\left|#1\right>\!\left<#1\right|}

\newcommand{\psl}{\mathcal{G}}

\DeclareMathOperator{\Tr}{\text{Tr}}
\newcommand{\deff}{\vcentcolon=}

\def\bf{\textbf}
\def\CC{\mathbb C}
\def\RR{\mathbb R}

\def\EE{\mathbb E}
\def\mcal{\mathcal}
\def\mathscr{\mscr}

\newcommand{\reff}[1]{Eq.~\eqref{#1}}  
\begin{document}
\title{Stochastic unraveling of positive quantum dynamics}   
\author{Matteo Caiaffa}
\email{matteo.caiaffa@strath.ac.uk}
\affiliation{SUPA and Department of Physics, University of Strathclyde, Glasgow G4 0NG, UK}
\author{Andrea Smirne}
\affiliation{Institute of Theoretical Physics, Universit{\"a}t Ulm, Albert-Einstein-Allee 11D-89069 Ulm, Germany}
\author{Angelo Bassi}
\affiliation{Department of Physics, University of Trieste, Strada Costiera 11, 34151 Trieste, Italy}
\affiliation{Istituto Nazionale di Fisica Nucleare, Trieste Section, Via Valerio 2, 34127 Trieste, Italy}
\begin{abstract}
Stochastic unravelings represent a useful tool to describe the
dynamics of open quantum systems, and standard methods, such as quantum state diffusion (QSD),
call for the complete positivity of the open-system dynamics.
Here, we present a generalization of QSD, which also applies to positive, but not completely positive
evolutions. The rate and the action of the diffusive processes involved in the unraveling
are obtained by applying a proper transformation to the operators which define the master equation. The unraveling 
is first defined for semigroup dynamics and then extended to a definite class of time-dependent generators.
We test our approach on a prototypical model for the description of exciton transfer,
keeping track of relevant phenomena, which are instead disregarded within the standard, completely positive framework.\\

\noindent
DOI: \href{https://doi.org/10.1103/PhysRevA.95.062101}{10.1103/PhysRevA.95.062101}
\end{abstract}
\maketitle
%
%
%
%
%
%
%
%
%
%
%
%
%
%

\noindent
\section{Introduction}%
The investigation of open quantum systems coupled to complex and possibly structured
environments has led to a renewed interest toward
the description of quantum dynamics beyond the paradigm of completely positive (CP) semigroups \cite{Breuer2002,carm2009,Rivas2012},
as fixed by the well-known Gorini-Kossakowski-Sudarshan-Lindblad (GKSL) master equation (ME) \cite{Gorini1976,Lindblad1976}.
The development of more general approaches has made possible to take into account memory effects and others phenomena
which are neglected within that framework; see for example the recent reviews \cite{Rivas2014,Breuer2016,Vega2016}. 

Mostly, the assumption to have a semigroup dynamics is relaxed, while one holds firm that the evolution
has to be given by CP maps. If there are no initial correlations between the system and the environment
and the initial state of the latter is fixed, the exact reduced dynamics, mathematically obtained via the partial trace on the environmental degrees of freedom, is indeed CP \cite{Breuer2002,carm2009,Rivas2012}. 
On the other hand, the partial trace can be hardly ever performed explicitely, even with powerful numerical techniques.
The restriction to CP maps becomes then questionable, not only when
initial correlations have to be considered \cite{Pechukas1994,Shaji2005,Modi2012,Buscemi2014}, but also when one uses an approximated description for specific open quantum systems at hand.
The weaker condition that the dynamics is positive (P) may be enough to guarantee the consistency of the predictions one is interested in.

In addition, when a ME is derived from some underlying microscopic model, CP is usually obtained by introducing some specific approximations, which, needless to say,
may overlook some relevant phenomena. 
As a paradigmatic example, in the weak coupling
regime one imposes (on top of the Born-Markov approximation) the secular approximation. The latter is justified when the free dynamics of the system
is much faster than its relaxation~\cite{Breuer2002}, which is not the case for 
several systems of interest.  
Non-secular non-CP evolutions, possibly still in the semigroup regime, are extensively used, e.g.,
to model transport phenomena in nanoscale biomolecular networks \cite{May2004,Ishizaki2009,Jeske2015,Oviedo2016}.

Certainly, CP evolutions possess several advantages, mainly due to the general mathematical results which allow for their full characterization,
such as the Kraus decomposition or the GKSL theorem itself \cite{Breuer2002}. Moreover, CP evolutions
have	 been equivalently formulated in terms of unravelings in the form of stochastic trajectories, being they with jumps \cite{Dalibard1992,Plenio1998} or continuous \cite{GisPer,rigo1997,WisDio,adler2001,Percival2002,Barchielli2009}.  
These methods yield a very powerful tool to simulate numerically open-system dynamics, as well as a deeper understanding of the different effects induced on the system
by the interaction with the environment, as in the theory of continuous measurement (see \cite{Wis96} and references therein). Also, unravelings of MEs play
 a role in the foundations of quantum mechanics, in connection with decoherent histories \cite{dio96,Brun00} and quantum state reduction theories \cite{grw,Bassi2003,Bassi2013,Bassi2013b}.

Here, we prove that a proper unraveling can be generally formulated also for P, not necessarily CP dynamics.
We focus in particular on a continuous form of the unraveling, the so-called quantum state diffusion (QSD) \cite{GisPer,rigo1997,WisDio,adler2001,Percival2002},
and we show how it can be directly extended to the more general case of P dynamics.
The role of the rates and Lindblad operators in the CP unraveling is replaced by the eigenvalues and eigenvectors
of a rate operator \cite{Dio86,Dio88,GisPer,WisDio}. Our approach includes not only semigroup dynamics,
but also a more general kind of evolutions; namely, P-divisible dynamics \cite{Vacchini2011,basbreu,Chruscinski2014,Bernardes2015,Liuzzo2016}, which has been recently taken into account within the context of the definition
of quantum Markovianity. In this way, we provide a significant class of open-system dynamics
with a useful tool to describe physical phenomena, which would be neglected within the usual CP framework.
This is explicitly shown by taking into account a model, which is of interest for the description of energy transfer in biomolecular networks \cite{Palmieri2009,Jeske2015,Oviedo2016}.

The rest of the paper is organized as follows. In Sect. \ref{sec:ucp}, we briefly recall the standard QSD unravelling of CP semigroups.
In Sect. \ref{sec:ups}, we introduce the QSD unravelling of P semigroups, which is then further extended to P-divisible maps in Sect. \ref{sec:upnm}.
In Sect. \ref{sec:exa}, we present two examples of P non CP dynamics, to which we apply our formalism; the first is a simple toy model for a qubit evolution, 
while the second is a significant model for the excitation transfer in dimeric systems.
Finally, the conclusions and future perspectives are given in Sect. \ref{sec:con}.

\section{Unraveling of CP semigroups}\label{sec:ucp}
Let us first briefly recall the standard results about (diffusive) unravelings of CP semigroups, as well as the relevant notation. 

We consider a finite dimensional quantum system, whose state $\rho$ is an element of the set $\mcal S(\CC^n)$ of positive trace-one operators on $\CC^n$. The dynamics is  described 
by a one-parameter family of linear maps $\left\{\Lambda_t\right\}_{t\geq 0}$, where $\Lambda_t:\mcal S(\CC^n)\rightarrow\mcal S(\CC^n)$ evolves the state $\rho$ at the initial time $t_0=0$,
into the state  $ \rho_t=\Lambda_t[\rho]$ at time $t$. 
These maps satisfy the semigroup property whenever
$
\Lambda_t \Lambda_s = \Lambda_{t+s},  \forall t,s \geq 0,
$
and in this case they can be expressed as $\Lambda_t = e^{t\psl}$ for some \emph{generator} $\psl$, so that
$\rho_t$ is fixed by the ME $d \rho_t/ dt = \psl[\rho_t]$.
The maps $\Lambda_t$ ensure the trace and hermiticity preservation
of the system's state $\rho_t$ if and only if the generator
$\psl$ can be written as \cite{Gorini1976} 
\begin{equation}\label{diag}
\psl[\rho]\deff - i \left[H, \rho\right]  + \sum_{j=1}^{n^2-1}c_j\left[ L_j\rho L_j^{\dagger}-\frac{1}{2}\left\lbrace L_j^{\dagger}L_j,\rho \right\rbrace \right],
\end{equation}
for some coefficients $c_j\in\RR$, linear operators  $L_j$ and an hermitian operator $H = H^{\dag}$. 
According to the GKSL theorem \cite{Gorini1976,Lindblad1976}, the maps $\Lambda_t$
generated by $\psl$ are CP if and only if $c_j \geq 0$ $\forall j$.

In addition to the GKSL theorem, another crucial feature of CP semigroups, further motivating their ubiquitous use to describe
open system's dynamics, is that they can be equivalently characterized via  \emph{unravelings}.
An unraveling consists of a 
stochastic dynamics for the pure states $|\psi\rangle$ of the system, which reproduces the ME under stochastic average.
Here, we focus on the case of a \emph{diffusive} unraveling, associated to a Stochastic Differential Equation (SDE) in the form \cite{GisPer,rigo1997,WisDio,adler2001,Percival2002,Barchielli2009}
\begin{equation}\label{sde1}
\ket{d\psi_t}=A_{\psi_t}\ket{\psi_t}dt+\sum_{k=1}^{m}B_{\psi_t,k}\ket{\psi_t}d\xi_{k,t},
\end{equation}
where
$A_{\psi_t},B_{\psi_t,k}$ are (possibly non-linear) operators and $\xi_{k,t}$ are independent
complex-valued Wiener processes, 
with
$
\EE [d\xi_{j,t}d\xi_{k,t}^{*}]=\delta_{jk}dt, \quad \EE \left[d\xi_{j,t}d\xi_{k,t}\right]=\EE \left[d\xi_{j,t}\right]=0, 
$
where $\EE$ denotes the statistical mean.
The resulting trajectories
in the Hilbert space are usually referred to as \emph{quantum trajectories}.
We always assume that the SDE preserves the norm of $\ket{\psi_t}$.

The connection with the 
statistical operator $\rho_t$ is obtained via the stochastic average $\EE$.
Given the stochastic projector $P_t\deff\proj{\psi_t}$ and its 
infinitesimal change $dP_t$ fixed by the It\^o formula,
$
dP_t=\projj{d\psi_t}{\psi_t}+\projj{\psi_t}{d\psi_t}+\proj{d\psi_t},
$
one says that Eq.~(\ref{sde1})
is an unraveling of Eq.~(\ref{diag})
when $
\psl[\rho_t]=\EE\left[dP_t/dt\right].
$
In general, there exist infinite unravelings for the same ME.
In the case of CP semigroups
the QSD unraveling is given by  \cite{GisPer,rigo1997,WisDio,adler2001,Percival2002} \reff{sde1}, with $m=n^2-1$ and
\begin{eqnarray}
A_{\psi_t} &&=-i H-\frac{1}{2}\sum_{j=1}^{n^2-1}c_j\big(L_j^\dag L_j-2\ell_{\psi, j}^*L_j+|\ell_{\psi, j}| ^2\big) \label{aucp}\\
B_{\psi_t, j} &&=\sqrt{c_j} \left(L_j-\ell_{\psi, j} \right), \label{bucp}
\end{eqnarray}
where $\ell_{\psi, j}\deff\bra\psi L_j\ket\psi$.


\section{Unraveling of P semigroups}\label{sec:ups}

\subsection{Constructive proof of the unraveling}

The previous approach can be extended to all the P, not necessarily CP semigroups,
(for Hilbert spaces of arbitrary finite dimension).
As long as we assume a semigroup evolution, P dynamics provide us with the largest class of dynamics that can have a norm-preserving unraveling
for any initial condition:
any state obtained via the statistical average is automatically positive, being the convex mixture of pure states.  
Later, we will see how the semigroup assumption can be replaced by a more general feature of the dynamics.

The unraveling of a P semigroup depends on the behaviour of a nonlinear operator, 
whose relevance for the unraveling of semigroups was already noticed in \cite{Dio86,Dio88,GisPer,WisDio}. Consider a generator as in \reff{diag}; for any normalized vector $\psi\in\CC^n$, 
we define the \emph{generalised transition rate operator} (GTRO)  as the linear combination~\cite{Dio86,Dio88,WisDio}
\begin{equation}\label{transition}
W_\psi\deff \sum_{j=1}^{n^2-1}c_j\big(L_j-\ell_{\psi, j}\big)\proj{\psi}\big(L_j-\ell_{\psi, j}\big)^\dag.
\end{equation}
The precise connection among the properties of this non-linear operator
and the unraveling of P semigroups traces back to the following
result, which is a direct consequence of a theorem by Kossakowski \cite{kossa,kossb}.

\noindent
\emph{Lemma 1.} The dynamical map $\Lambda_t = e^{t \psl}$ is P if and only if, for any normalized vector $\psi\in\CC^n$, $W_\psi$ is a positive semi-definite operator.


\noindent
\emph{Proof.} 
As noticed in \cite{basbreu}, the aforementioned Kossakowski's theorem \cite{kossa,kossb} can be rephrased as follows: 
given any orthonormal basis $\left\{\ket{u_{i}}\right\}_{i=1, \ldots n}$, then
\begin{equation}\label{bas}
(\rho\geq 0\Rightarrow\Lambda_t[\rho]\geq 0)\Leftrightarrow \sum_{j=1}^{n^2-1}c_j\left|\bra{u_{i}}L_j\ket{u_{i'}}\right|^2\geq 0,
\end{equation}
for any couple $i\neq i'$. 

Let us consider two arbitrary states $\ket{\psi},\ket{\varphi}$. Write 
\begin{equation}\label{var}
\ket\varphi=a\ket{\psi}+b\ket{\psi_\perp},
\end{equation}
where the two vectors on the r.h.s. are the components of $\ket{\varphi}$, which are parallel and perpendicular to $\ket{\psi}$, respectively. Notice the relations
\begin{align}
\bra\psi (L_j-\ell_{\psi, j})\ket{\psi}&=0,\label{uno1}\\
\bra\psi (L_j-\ell_{\psi, j})\ket{\psi_\perp}&=\bra\psi L_j\ket{\psi_\perp}\label{due2}.
\end{align}
Then, using \reff{uno1} and \reff{due2}, we obtain for any $\psi$ the equivalence
\begin{equation}\label{equivalence}
\bra\varphi W_\psi\ket\varphi= |b|^2\sum_{j=1}^{n^2-1}c_j\left|\bra{\psi} L_j\ket{\psi_\perp}\right|^2 \quad \forall \ket{\psi},  \ket{\varphi}.
\end{equation}
Given the equation above, the proof of the Lemma is straightforward. On the one hand, the positivity of $W_\psi$ for any $\ket{\psi}$ implies 
the positivity of the r.h.s. of \reff{bas} for any couple of orthogonal elements of any given basis (just set $\ket{\psi} = \ket{u_{i}} $ and 
$\ket{\psi_{\perp}} = \ket{u_{i'}} $), from which the positivity of the semigroup follows. 
One the other hand, if $\Lambda_t$ is P, and hence the r.h.s. of \reff{bas} is positive,
the non-negativity of the r.h.s. of \reff{equivalence} for any $\ket{\varphi}$ and $\ket{\psi}$, therefore the positive semidefiniteness of $W_\psi$ for any $\ket{\psi}$, directly follows from 
setting $\ket{u_{i}} = \ket{\psi}$ and using the decomposition of $\ket{\psi_{\perp}}$
on the elements of the basis $\ket{u_{i'}}$ orthogonal to $\ket{u_{i}}$, i.e. with $i\neq i'$.\qed

This result will be the building block of the construction of our unraveling of P semigroups. 
Some diffusive unravellings which can be applied beyond CP semigroups already appeared in the literature, for a qubit in \cite{Gisin1990}
and for any finite dimensional system in \cite{Dio88}. Nevertheless,
let us stress how the definite connection between the possibility to formulate an unraveling and the positivity of the corresponding semigroup dynamics
for any finite dimension was missing until now; see also the recent discussion in \cite{Dio2016}.

Proceeding further, Lemma 1 implies that when we have a semigroup of P maps and we consider the linear operator
$W_\psi$ for any fixed $\psi$,
its eigenvalues $\lambda_{\psi, i}$ ($i=0, \ldots, n-1$) are non-negative,
where $\lambda_{\psi, 0} = 0$ corresponds to the eigenvector $\ket{\psi}$,
so that we can write the spectral decomposition as
\begin{equation}\label{eigen}
W_\psi=\sum_{i=1}^{n-1}\lambda_{\psi, i}\proj{\phi_{\psi, i}}=\sum_{i=1}^{n-1}\lambda_{\psi, i}\big(V_{\psi, i}\proj{\psi}V_{\psi, i}^\dag\big),
\end{equation} 
with $\lambda_{\psi, i} \geq 0$ and $\proj{\phi_{\psi, i}}$ the corresponding orthogonal projectors, satisfying
$\braket{\phi_{\psi, i}}{\psi} = 0$. 
The second equivalence in \reff{eigen} is trivially justified by defining $V_{\psi, i}=\projj{\phi_{\psi, i}}{\psi}$, which will also provide us with a clear
physical interpretation of the unraveling.

Now, by using It\^o calculus, it is readily verified that \reff{sde1} yields the following SDE for $P_t=\proj{\psi_t}$:
\begin{eqnarray}\label{me}
dP_t&=&\left(A_{\psi_t} P_t +P_tA_{\psi_t}^\dag+\sum_{k=1}^{m}B_{\psi_t,k}P_t B_{\psi_t,k}^\dag\right)dt \\
&+&\sum_{k=1}^{m}(B_{\psi_t,k}P_td\xi_{k,t}+P_tB_{\psi_t,k}^\dag d\xi_{k,t}^*). \nonumber
\end{eqnarray}
In addition, since we want the SDE to be an unraveling of the ME fixed by $\psl$ at any time $t$, we are assuming, 
in particular, that this is the case at time $t=0$, i.e. $\EE\left[\left.dP_t/dt\right|_{t=0}\right] = \psl[\rho_0]$.
From this relation, along with \reff{me},
it follows that the noise term $\sum_{k=1}^{m}B_{\psi,k}P B_{\psi,k}^\dag$ is given by the component of $\psl\big[P\big]$ orthogonal to $\ket{\psi}$, i.e.
\begin{align}
\sum_{k=1}^m B_{\psi,k}PB_{\psi,k}^\dag&=\big(I-P\big)\psl\big[P\big]\big(I-P\big).\label{necessary}
\end{align}

The last statement, which was first shown in \cite{GisPer}, can be easily re-derived consistently with our notation, as shown in the the following.
First, let us take the expectation of \reff{me} for a deterministic initial condition, $\ket{\psi_0} =: \ket{\psi}$ so that $\rho_0 = P_0 =: P$; 
since $\EE\left[\left.dP_t/dt\right|_{t=0}\right] = \psl[\rho_0]$, we get
\begin{equation}\label{first}
A_\psi P +PA_\psi^\dag
+\sum_{k=1}^{m}B_{\psi,k}P B_{\psi,k}^\dag
=\psl\big[P\big].
\end{equation}
The SDE in \reff{sde1} preserves the norm of the state vector only if 
\begin{equation}\label{para1}
\bra\psi B_{\psi,k}\ket\psi=0 \qquad \forall \psi, k.
\end{equation}
Then, if we denote by $\ket{\psi_{\perp}}$ a vector orthogonal to $\ket{\psi}$, the norm constraint translates into $B_{\psi,k}\ket\psi=\ket{\psi_{\perp}}$. 
In other words, the noise operators must produce orthogonal changes to the state vector they act upon.
For any fixed $\ket{\psi}$ this condition implies
\begin{equation}
P\left(\sum_{k=1}^m B_{\psi,k}PB_{\psi,k}^\dag\right) P = 0;\label{perp}
\end{equation}
on the other hand, 
\begin{equation}
(\mathbb{I}-P)\left(A_\psi P+PA_\psi^\dag\right)(\mathbb{I}-P) = 0,\label{par}
\end{equation}
so that by projecting \reff{first} on the subspace orthogonal to $\ket{\psi}$, 
\reff{perp} together with \reff{par} prove the validity of \reff{necessary}. 

Now, with the help of simple algebra, \reff{necessary} reduces to
\begin{equation}
\sum_{k=1}^m B_{\psi,k}PB_{\psi,k}^\dag=W_\psi.\label{neces2}
\end{equation}
We can conclude that  \reff{neces2} has to be satisfied by \emph{all}  possible (norm preserving) unravelings, as in \reff{sde1}, of the ME fixed by \reff{diag}.

Moreover, \reff{neces2}, along with \reff{first}, imply that the action of the drift operator $A_\psi$ on the state $\ket{\psi}$ is determined by $W_{\psi}$ and the generator $\psl$ via
\begin{equation}\label{alp}
A_\psi P +PA_\psi^\dag=\psl\big[P\big]- W_\psi.
\end{equation}
This means that $A_\psi$ can be set independently from the specific solution of \reff{neces2} for the $B_{\psi,k}$,
and, in particular, $A_\psi$ is still fixed by \reff{aucp}. To see the \emph{uniqueness} of such a choice, 
notice that \reff{first}, along with \reff{neces2}, leads to \reff{alp}, as can be easily checked by writing explicitly $\psl[P]$ and $W_{\psi}$ via, respectively, \reff{diag} and \reff{transition}. Let us emphasize that, indeed, this is not the only non-linear operator satisfying \reff{alp},
but any other solution $\tilde{A}_{\psi}$ would act on the state $\psi$
exactly in the same way, such that
\begin{equation}\label{apr}
\tilde{A}_{\psi_t}\ket{\psi_t} = A_{\psi_t} \ket{\psi_t};
\end{equation}
in other terms, it would lead exactly to the same unraveling, see \reff{sde1}:
in this regard, the choice of $A_{\psi}$, for fixed $B_{\psi, k}$ and noise $\xi_{k, t}$, is \emph{unique}.
To prove the validity of \reff{apr}, consider two different solutions, $A_{\psi}$ and $\tilde{A}_{\psi}$, to \reff{alp}.
Then, $A_\psi P +PA_\psi^\dag = \tilde{A}_\psi P +P\tilde{A}_\psi^\dag$. Hence, for any state $\ket{\psi_\perp}$
orthogonal to $\ket{\psi}$, one has
$$
\bra{\psi_\perp}A_{\psi} \ket{\psi} = \bra{\psi_\perp}\tilde{A}_{\psi} \ket{\psi}
$$
and, analogously for the parallel component,
$$
\mbox{Re}[\bra{\psi}A_{\psi} \ket{\psi}] =\mbox{Re}[\bra{\psi}\tilde{A}_{\psi} \ket{\psi}].
$$
In principle, $A_{\psi_t} \ket{\psi_t}$ and $\tilde{A}_{\psi_t} \ket{\psi_t}$ could differ by a purely imaginary component parallel to $\ket{\psi_t}$;
but it is then easy to see \cite{Bassi2013b} that such a difference corresponds simply to an irrelevant global phase applied to $\ket{\psi_t}$.


All in all, to define a proper unraveling of a P semigroup, we are simply left with formulating a solution of \reff{neces2}.
A natural choice is given by
the spectral decomposition of $W_{\psi}$, which, by virtue of the positivity of the semigroup and then Lemma 1, 
is characterized by the non-negative eigenvalues $\lambda_{\psi, k}$.
Hence, let us set $m=n-1$ and
\begin{equation}\label{hb}
B_{\psi,k}=\sqrt{\lambda_{\psi,k}}V_{\psi,k}.
\end{equation} 
It is then easy to see that $B_{\psi,k}$ as in \reff{hb} satisfies \reff{neces2} and, along with $A_\psi$ as in \reff{aucp}, defines a SDE as in \reff{sde1}
which provides us with a proper unraveling of the P semigroup generated by \reff{diag}. We thus arrived to the wanted result: Eqs.~(\ref{aucp}) and (\ref{hb})
generalize the QSD unraveling of CP semigroups to the case of P, not necessarily CP,
semigroups. Note that, for $c_j\geq0$ a solution to \reff{neces2} is directly provided by $m=n^2-1$ and
$B_{\psi,k} =\sqrt{c_{k}}(L_{k}-\ell_{\psi,k})$, so that one recovers \reff{bucp}.
On the other hand, when some $c_j$ takes on a negative value (as in the P non CP case), a solution of \reff{neces2} as in \reff{bucp} would give a set of SDEs as in \reff{sde1} which are not consistent with the average dynamics: by deriving the stochastic MEs through $P_t\deff\proj{\psi_t}$, one would get the positive coefficients $|c_j|$.\\  

The crucial point for extending the unraveling to every P semigroup
 is the observation that the role of the rates $c_j$ can be replaced by the eigenvalues $\lambda_{\psi,i}$ in the spectral decomposition (\ref{eigen}) of $W_\psi$, whose positivity is ensured by Lemma 1. 
 Accordingly, the operators $V_{\psi, i}$ replace the Lindblad operators $L_j$ (of course, $\bra\psi V_{\psi, i}\ket\psi = 0$). 
The physical meaning of the unraveling here defined is hence quite clear: the eigenvalues and eigenvectors
of the GTRO set, respectively, the strength of the diffusive processes
and how they act on the elements of the Hilbert space. In particular, $V_{\psi_t,i}$ maps the stochastic state at time $t$, $\ket{\psi_t}$,
into the state $\ket{\phi_{\psi_t, i}}$, which appears in the spectral decomposition of $W_{\psi_t}$ and is orthogonal to $\ket{\psi_t}$.
To deal with P, but not CP semigroups we exploit the diagonalization of the GTRO, while in the CP case 
the coefficients and operators in the Lindblad generator directly fix the quantum trajectories; the difference between 
the two cases will be illustrated later for a specific example.

\subsection{CP and norm preservation}%
As a further remark, we note how the previous results imply that, indeed, the requirement of getting a closed ME from a diffusive norm preserving SDE does not imply, by itself, CP. 
This was shown by direct counterexample in \cite{Diosi2014} (see also Sect. \ref{sec:uncp}),
and it can be easily related to the lack of norm preservation of the ME unraveling extended to an arbitrary ancilla.

To see this, let us recall that a linear map $\Lambda:\mcal S(\CC^n)\to\mcal S(\CC^n)$ is CP if and only if the map 
$\Lambda\otimes\mathbb{I}:\mathcal{S}(\CC^n\otimes\CC^n)\to\mathcal{S}(\CC^n\otimes\CC^n)$ is P. Let $\mathcal{G}$ 
and $\mathcal{G}'$ be the generators of $\Lambda$ and $\Lambda\otimes\mathbb{I}$, respectively. 
Assume $\Lambda$ (at least) P, and call ${d\psi}$ the norm preserving unraveling of its generator. Moreover, we define ${d\psi}'$ 
to be a particular extension of the original SDE to an enlarged Hilbert space, such that it reproduces, on average, $\mathcal{G}'$. 
Then, we are led to the following diagram
\begin{equation}\label{dgr}
\xymatrix{
   d\psi \ar@{->}[r]^{} \ar@{<->}[d]_{} & d\psi' \ar@{<->}[d]_{} \\
   \mathcal{G}[\rho] \ar@{->}[r]^{} & \mathcal{G}'[\rho'] 
  }
\end{equation}

\noindent
where the vertical arrows represent the operations of unraveling and taking the stochastic average, and 
the horizontal ones stand for tensoring with auxiliary operators, in such a way that the diagram commutes.

Now let us assume that $\Lambda_t$ is not CP. Then, there exist $\rho'\in\mathcal{S}(\CC^n\otimes\CC^n)$ such that  
$\bar{\rho}=(\Lambda_t\otimes\mathbb{I})[\rho']=\EE[P_t']$, where $P'_t=\ket{\psi_t'}\bra{\psi_t'}$, is not a proper quantum state, i.e. $\bar{\rho}$ is
 either not positive, not trace one, or both. However, any operator obtained via stochastic average is positive, being the convex combination of the positive operators 
 $P'_t$. Then, if $\Lambda\otimes\mathbb{I}$ is not P, it must be the case that $\Tr(\EE[P'_t])=\EE[\Tr(P'_t)]\neq 1$, i.e. $\ket{d\psi'}$ does not preserve
 the norm of all state vectors. In summary, under the hypothesis that diagram (\ref{dgr}) commutes, asking that the extended SDE be norm preserving is a sufficient condition for the CP of $\Lambda$.

\section{Unraveling of P-divisible dynamics and relation with Markovianity}\label{sec:upnm}

Our approach can be straightforwardly generalized
to a much wider class of dynamics, which goes beyond the class that can be treated
via the usual unravelings for CP maps. We consider now evolutions where the coefficients, and possibly the operators, in the ME depend on time.
This allows to describe several situations of interest, where the semigroup approximation cannot be used,
because time inhomogeneous and non-Markovian effects become relevant \cite{Rivas2014,Breuer2016,Vega2016}.

Consider a \emph{time-dependent} generator $\psl_t$. Once again, trace and hermicity preservation constrain it to have the form 
as in \reff{diag}, at any time $t$, i.e., one has
\begin{eqnarray}\label{diagt}
\psl[\rho]& \deff & - i \left[H(t), \rho\right]  \\
&&+ \sum_{j=1}^{n^2-1}c_j(t)\left[ L_j(t)\rho L_j(t)^{\dagger}-\frac{1}{2}\left\lbrace L_j(t)^{\dagger}L_j(t),\rho \right\rbrace \right],\nonumber
\end{eqnarray}
where now we have a time-dependent hamiltonian, as well as time-dependent rates and Lindblad operators.
Note that the CP of the dynamics does not imply the positivity of the coefficients (since, indeed, the GKLS theorem does not apply).
The most general conditions to guarantee CP, not to mention P, of the resulting dynamical maps
$\Lambda_t = \mathcal{T} \exp\left(\int_0^t \psl_s d s\right)$ (with $\mathcal{T}$ the time-ordering operator)
are actually not known. Nevertheless, the positivity in time of the coefficients, $c_j(t) \geq0$,
guarantees that
the dynamics is CP and can be decomposed into intermediate CP maps \cite{Laine2010,Rivas2012}:
for any $t \geq s \geq 0$, there is a CP map $\Lambda_{t,s}$ such that 
\begin{equation}\label{eq:dec}
\Lambda_t = \Lambda_{t,s} \circ \Lambda_s;
\end{equation}
in this case the dynamics is said to be \emph{CP-divisible} and this property
has been identified with the \emph{Markovianity} of the quantum dynamics in \cite{Rivas2010}.
Note that the positivity of the coefficients allows one to extend the QSD unraveling of CP semigroups
to this case: one has simply to replace $c_j \rightarrow c_j(t)$, $L_j \rightarrow L_j(t)$ and  $H\rightarrow H(t)$
in Eqs.~(\ref{aucp}) and (\ref{bucp}).


The unraveling defined via Eqs.~(\ref{aucp}) and (\ref{hb}) can also be extended
to ME with time-dependent coefficients, which need not be positive
functions of time. Consider any ME leading to a dynamics which, instead of being CP-divisible,
is \emph{P-divisible}, which means that the decomposition in \reff{eq:dec} still applies,
but now we make the weaker requirement that the maps $\Lambda_{t,s}$ are P \cite{Vacchini2011};
this property, in turn, has been identified with quantum Markovianity in \cite{basbreu}.
The construction presented before can be immediately generalized to this situation,
since the equivalence in Lemma 1 still applies. Indeed, the extension of Lemma 1 to the case of P-divisible dynamics directly follows from the analogous extension of the theorem by Kossakowski,
pointed out in \cite{basbreu}. To see this, let us consider a ME as in \reff{diagt}.
The resulting dynamical map $\Lambda_t$ is P and can be decomposed
via \reff{eq:dec} with P $\Lambda_{t,s}$ if and only if
\begin{equation}\label{bas2}
\sum_{j=1}^{n^2-1}c_j(t)\left|\bra{u_{i}}L_j(t)\ket{u_{i'}}\right|^2\geq 0,
\end{equation}
for any couple $i\neq i'$ \cite{basbreu}. But then, similarly to the proof for the semigroup case one can show 
that the latter condition is equivalent to the positivity of $W_{\psi}$, defined as in \reff{transition},
with the replacements $c_j \rightarrow c_j(t)$ and $L_j \rightarrow L_j(t)$, so that
Eqs.~(\ref{aucp}) and (\ref{hb}), with the proper introduction of time-dependence, define a valid unraveling of a generic P-divisible ME.

Of course, there are several open-system dynamics which are not P-divisible and, therefore, cannot be unravelled via our approach,
but where other diffusive \cite{Diosi1997,Gambetta2002} or jump \cite{Piilo2008} techniques can be exploited. On the other hand,
our approach 
yields a direct generalization of the construction for the semigroup case,
without calling for hierarchical equations, nor for correlations between different trajectories, 
which are instead usually required by the above-mentioned techniques. 
 
As a final remark, we note that Eqs.~(\ref{sde1}), (\ref{aucp}) and (\ref{hb}) comprise the most general (Markovian) 
dynamics of collapse models~\cite{grw,Bassi2003,Bassi2013,Bassi2013b}. Here, it suffices to say that collapse models consist in a modification of
the Schr{\"o}dinger equation with the addition of non-linear stochastic terms, which ensure the localization of the wave function. 
The dynamics of collapse models is usually defined as a diffusion process in the Hilbert space (given by a SDE as in \reff{sde1}), 
although piece-wise evolutions involving jumps processes are also possible \cite{grw}. If we limit to a dynamics as in \reff{sde1}, the requirement of getting a closed linear average description,
which is physically motivated by the request of no-superluminal-signaling \cite{Gisin89,GisinRigo}, is not enough to guarantee the CP.
As said, this traces back to the possible lack of norm preservation
of the SDE trivially extended to an arbitrary ancilla.
Such an extension, indeed, would be rather unmotivated for collapse models,
since the collapsing field would act also on the ancilla, possibly in a non-local way:
in this context, the CP of the ensemble dynamics is an extra assumption, not 
emerging from fundamental requirements.

\section{Examples}\label{sec:exa}

\subsection{Unraveling of a non-CP qubit ME}\label{sec:uncp}
Here, we consider the unravelling of the non-CP semigroup which was first derived in \cite{Gisin1990,Diosi2014}.
Although the physical relevance of the model is not clear, it is the first example of an unravelling of a P, but non CP semigroup
and it was thus used in \cite{Diosi2014} to prove that the CP of the average dynamics is not guaranteed by the existence
of a Markovian unravelling. We will show now how such a result can be straightforwardly re-derived and further clarified using our method.
 
Hence, consider the non-CP semigroup acting on $\mcal S(\CC^2)$ and generated by
\begin{equation}\label{dios1}
\frac{d\rho_t}{dt}=\sum_{j=1}^3c_j(\sigma_j\rho_t\sigma_j-\rho_t), \ \ c_1=c_2=-c_3=1,
\end{equation}
where $\sigma_j$ are the usual Pauli matrices $\sigma_1\equiv\sigma_x$, $\sigma_2\equiv\sigma_y$ and $\sigma_3\equiv\sigma_z$. 
The  GTRO associated to \reff{dios1} is
\begin{equation}\label{gtro_ex}
W_\psi=\sum_{j=1}^3c_j(\sigma_j-s_j)\proj{\psi}(\sigma_j-s_j)
\end{equation}
with $s_j\deff\bra{\psi}\sigma_j\ket{\psi}$, and it has spectral decomposition 
\begin{equation*}
W_\psi=\lambda_1\proj{\psi}+\lambda_2\proj{\psi_\perp},
\end{equation*}
where the eigenvalues are $\lambda_1=0$ and $\lambda_2=2\,s_3^2$, while the eigenvectors are $\ket{\psi}$ and $\ket{\psi_\perp}$ orthogonal to $\ket\psi$. The first eigenvalue and 
eigenvector can be easily found by noticing that $W_\psi\ket{\psi}=0$, as \reff{uno1} ensures (see also the discussion before \reff{eigen}). Then, we are left with verifying that  
\begin{align}
(W_\psi-2\,s_3^2)\ket{\psi_\perp}&=\sum_{j=1}^3c_j\bigg(\sigma_jP\sigma_j
-s_jP\sigma_j-2s_3^2\bigg)\ket{\psi_\perp}\nonumber\\
&=0\label{specdec}. 
\end{align}
To show that, we project \reff{specdec} on the basis vectors $\bra{\psi}$ and $\bra{\psi_\perp}$, respectively. Since for any $j$
\begin{align*}
\bra{\psi}\sigma_jP\sigma_j\ket{\psi_\perp}&=
\bra{\psi}\sigma_j\proj{\psi}\sigma_j\ket{\psi_\perp}\\
&=\bra{\psi}s_j\proj{\psi}\sigma_j\ket{\psi_\perp}\\
&=\bra{\psi}s_jP\sigma_j\ket{\psi_\perp},
\end{align*}
we have
\begin{align}
\bra{\psi}(W_\psi-2\,s_3^2)\ket{\psi_\perp}&=\sum_{j=1}^3c_j\bra{\psi}(\sigma_jP\sigma_j-s_jP\sigma_j)\ket{\psi_\perp}\nonumber\\
&=0\label{pp}
\end{align}
On the other hand,
since $\sum_{j=1}^3s_j^2=1$ and $r_j=|\bra{\psi}\sigma_j \ket{\psi_{\perp}}|^2=1-s_j^2$,
we have
\begin{align}
\bra{\psi_\perp}(W_\psi-2\,s_3^2)\ket{\psi_\perp}&=\sum_{j=1}^3c_j\bra{\psi_\perp}
\sigma_j P\sigma_j\ket{\psi_\perp}-2s_3^2\nonumber\\
&=\sum_jc_jr_j-2s_3^2=0.\label{pr}
\end{align}
 \reff{pp} together with \reff{pr} prove \reff{specdec}, which implies $W_\psi=2s_3^2\proj{\psi_\perp}$.

Then, according to \reff{eigen} and (\ref{aucp}), the noise and the drift terms which define a unraveling of \reff{dios1} are given, for any $\ket{\psi}$, by
\begin{align*}
B_{\psi_t}&=\sqrt{2}s_3\projj{\psi_{t\perp}}{\psi_t}\\
A_{\psi_t}&=-iH-\frac{1}{2}\sum_{j=1}^3c_j(\sigma_j-s_j)^2,
\end{align*}
so that 
\begin{equation}\label{unrvl}
\ket{d\psi_t}=-iH-\frac{1}{2}\sum_{j=1}^3(\sigma_j-s_j)^2\ket{\psi_t} dt+\sqrt{2}\,s_3\ket{\psi_{t\perp}} d\xi_{j,t}.
\end{equation}
Clearly, one can verify that \reff{unrvl} is norm preserving ($\braket{\psi}{d\psi}+\braket{d\psi}{\psi}+\braket{d\psi}{d\psi}=0$) and generates, on average, the ME
in \reff{dios1}.

\subsection{The Bloch-Redfield equation for a dimer system}%
As a specific, physically relevant application of our approach, we consider an example given by a simple description of a dimer system, 
which nevertheless represents a useful model to investigate exciton transfer, for example in biomolecular complexes~\cite{Palmieri2009,Jeske2015,Oviedo2016}.

The state of the excitation is associated with a three-level system: two levels for the excitation being in one or the other site,
and one level for the absence of excitation. The most relevant sources of noise are the pure dephasing and the recombination process. Using a perturbative approach (e.g., projection
operator techniques) up to second order and the Born-Markov approximation, one gets the Bloch-Redfield equation \cite{Breuer2002}.
This equation usually does not guarantee the positivity of the evolution and it is
then further approximated by a Lindblad equation, which even ensures that the dynamics is CP. The Lindblad equation is obtained via the secular approximation (SA),
which essentially neglects all the terms coupling population and coherences of the system.
However, this approximation is not always justified from a physical point of view, as it calls for a large difference
in the time scales of the free evolution and the dissipative relaxation of the system.
To overcome this difficulty and retain all the relevant phenomena in the dimer evolution, yet in a semigroup description of the dynamics,
a partial SA was introduced in \cite{Jeske2015}. The latter discards only 
some terms which couple population and coherences, while it preserves the most relevant ones. The resulting ME 
implies a P, but in general not CP evolution. Hence, it provides us with a natural benchmark to test our method.

The ME both after the full and the partial SA can be written as \cite{Jeske2015}
\begin{equation}\label{eq:eqje}
\dot{\rho}_{ij} (t) = \sum^3_{k l=1} \mathcal{R}^{\chi}_{ij;kl} \rho_{kl}(t),
\end{equation}
where $\rho_{kl}(t) = \bra{k} \rho(t) \ket{l}$. We will use the notation $\chi=CP$ for the full SA, while $\chi = P$
for the partial SA.
In order to write the ME (\ref{eq:eqje}) into the Lindblad form as in Eq.~(\ref{diag}), let us report the explicit expression of the coefficients $\mathcal{R}^{\chi}_{ij;kl}$. In the case of the partial SA they are given by [see Eq.~(11) in \cite{Jeske2015}]
\begin{eqnarray}
\mathcal{R}^{\mbox{{\footnotesize{P}}}}_{11,11}&=&\mathcal{R}^{\mbox{{\footnotesize{P}}}}_{22,22}=\mathcal{R}^{\mbox{{\footnotesize{P}}}}_{33,33}/2=-4 \nonumber\\
\mathcal{R}^{\mbox{{\footnotesize{P}}}}_{11,33}&=&\mathcal{R}^{\mbox{{\footnotesize{P}}}}_{33,11}=\mathcal{R}^{\mbox{{\footnotesize{P}}}}_{22,33}=\mathcal{R}^{\mbox{{\footnotesize{P}}}}_{33,22}=4 \nonumber\\
\mathcal{R}^{\mbox{{\footnotesize{P}}}}_{11,12}&=&\mathcal{R}^{\mbox{{\footnotesize{P}}}}_{22,21}=\mathcal{R}^{\mbox{{\footnotesize{P}}}}_{31,32}=\mathcal{R}^{\mbox{{\footnotesize{P}}}}_{32,31} = -71 i \nonumber\\
\mathcal{R}^{\mbox{{\footnotesize{P}}}}_{22,12}&=&\mathcal{R}^{\mbox{{\footnotesize{P}}}}_{11,21}=\mathcal{R}^{\mbox{{\footnotesize{P}}}}_{13,23}=\mathcal{R}^{\mbox{{\footnotesize{P}}}}_{23,13} =  71 i \nonumber\\
\mathcal{R}^{\mbox{{\footnotesize{P}}}}_{21,11}&=&\mathcal{R}^{\mbox{{\footnotesize{P}}}*}_{12,11}=-\mathcal{R}^{\mbox{{\footnotesize{P}}}*}_{12,22}=-\mathcal{R}^{\mbox{{\footnotesize{P}}}}_{21,22}=-1+71 i \nonumber\\
\mathcal{R}^{\mbox{{\footnotesize{P}}}}_{12,12}&=&\mathcal{R}^{\mbox{{\footnotesize{P}}}*}_{21,21} = -8-46 i \nonumber\\
\mathcal{R}^{\mbox{{\footnotesize{P}}}}_{13,13}&=&\mathcal{R}^{\mbox{{\footnotesize{P}}}*}_{31,31} = -9+12210 i \nonumber\\
\mathcal{R}^{\mbox{{\footnotesize{P}}}}_{23,23}&=&\mathcal{R}^{\mbox{{\footnotesize{P}}}*}_{32,32} = -9+12256 i,\label{eq:rp}
\end{eqnarray}
and all the other coefficients are equal to 0; as one can directly check, this provides us with a P, but not CP evolution.
On the other hand, as widely discussed in \cite{Jeske2015},
a CP evolution is obtained with a full SA, which means that the
terms coupling populations and coherences are set to 0:
\begin{eqnarray}
\mathcal{R}^{\mbox{{\footnotesize{CP}}}}_{11,12}&=&\mathcal{R}^{\mbox{{\footnotesize{CP}}}}_{22,21}=\mathcal{R}^{\mbox{{\footnotesize{CP}}}}_{31,32}=\mathcal{R}^{\mbox{{\footnotesize{CP}}}}_{32,31} =0 \nonumber\\
\mathcal{R}^{\mbox{{\footnotesize{CP}}}}_{22,12}&=&\mathcal{R}^{\mbox{{\footnotesize{CP}}}}_{11,21}=\mathcal{R}^{\mbox{{\footnotesize{CP}}}}_{13,23}=\mathcal{R}^{\mbox{{\footnotesize{CP}}}}_{23,13} = 0 \nonumber\\
\mathcal{R}^{\mbox{{\footnotesize{CP}}}}_{21,11}&=&\mathcal{R}^{\mbox{{\footnotesize{CP}}}}_{12,11}=\mathcal{R}^{\mbox{{\footnotesize{CP}}}}_{12,22}=\mathcal{R}^{\mbox{{\footnotesize{CP}}}}_{21,22} =0, \label{eq:rcp}
\end{eqnarray}
while all the other coefficients in \reff{eq:rp} are not changed. 
The parameters appearing in the two MEs express, in units of $\text{cm}^{-1}$, the effect
of dephasing and recombination noise, which are modeled as a spatially uncorrelated noise with an Ohmic spectrum \cite{Jeske2015}, as well as the Hamiltonian part of the dynamics.
In particular, note that the energy difference between the ground state and the two excitonic states is two or three orders of magnitude larger than any other relevant parameter in the free Hamiltonian \cite{Hoyer2010},
which explains the appearance of imaginary components in the ME parameters which are much bigger than the other values.

As described in Sect. \ref{sec:ups}, to define our unraveling we first need 
to write down the Lindblad form of the ME, which can be readily obtained following \cite{Gorini1976}. 
First, note that the generator $\mathcal{G}$ can be directly reconstructed via the coefficients in Eq.~(\ref{eq:eqje}),
since
\begin{equation}
\mathcal{R}^{\chi}_{ij;kl} = \bra{i}\mathcal{G}^{\chi}\left[\ketbra{k}{l}\right]\ket{j}.
\end{equation}
Then, consider the basis of operators on $\CC^3$ given by $\left\{\tau_i\right\}_{i=0,\ldots8}$,
with $\tau_0=\mathbb{1}/\sqrt{3}$, while the $\tau_i$s with $i=1,\ldots 8$
are the Gell-Mann matrices over $\sqrt{2}$ (to guarantee the normalization with respect to the Hilbert-Schmidt scalar product).
Hence, the so-called non-diagonal form of the generator $\mathcal{G}$ is given by
\begin{equation}\label{ndiag}
\psl[\rho]\deff - i \left[H, \rho\right]  + \sum_{ij=1}^{8}d_{ij}\left[ \tau_i\rho \tau_j^{\dagger}-\frac{1}{2}\left\lbrace \tau_j^{\dagger}\tau_i,\rho \right\rbrace \right],
\end{equation}
with
\begin{eqnarray}
H &=& \frac{1}{2i} \left(\tau^{\dag}-\tau\right), \quad \tau=\frac{1}{3}\sum^8_{i=1, k=0} \mbox{Tr}\left\{\tau_k \tau_i \mathcal{G}[\tau_k]\right\} \tau_i \nonumber\\
d_{i j}&=& \sum^8_{k=0}\mbox{Tr}\left\{\tau_j \tau_k \tau_i \mathcal{G}[\tau_k]\right\} \quad i,j=1,\ldots8.
\end{eqnarray}
The matrix of coefficients $d_{ij}$ is Hermitian, as the dynamics is Hermiticity preserving;
so there is a unitary matrix $U$, with elements $U_{ij}$, which diagonalizes it.
The resulting coefficients of the diagonal matrix are just the coefficients $c_j$ appearing in the diagonal form of $\mathcal{G}$
in Eq.~(\ref{diag}), and the matrix $U$ also defines the corresponding Lindblad operators $L_j$: explicitly one has
\begin{eqnarray}
c_j &=& \sum_{kk'=1}^8 U^{*}_{kj} d_{kk'}U_{k'j} \nonumber\\
L_j &=&\sum_{i=1}^8 U_{ij}\tau_i.
\end{eqnarray}
For the generator $\mathcal{G}^{P}$ fixed by Eq.~(\ref{eq:rp}) we get
the coefficients
\begin{eqnarray}\label{35}
c_1 &=& 2+\sqrt{5}, \quad c_2=c_3=c_4=c_5=4,\\
c_6&=&\frac{1}{3}\left(4+\sqrt{19}\right), \quad c_7 = 2-\sqrt{5} \quad c_8 =\frac{1}{3}\left(4-\sqrt{19}\right), \nonumber
\end{eqnarray}
where, note, the last two are negative, thus witnessing the non CP of the resulting semigroup dynamics.
The corresponding (canonical) Lindblad operators are
\begin{eqnarray}\label{36}
L_1 &=& -f_{1,-} \tau_1+f_{1,+} \tau_3 \quad L_7 = f_{1,+} \tau_1+f_{1,-} \tau_3   \nonumber \\
L_2&=&\tau_4 \quad L_3= \tau_5 \quad L_4=\tau_6 \quad L_5=\tau_7 \nonumber\\
L_6&=&i f_{2,-} \tau_2+f_{2,+} \tau_8 \quad L_8 = -i f_{2,+} \tau_2 +f_{2,-}\tau_8\nonumber\\
f_{1,\pm} &=& \sqrt{\frac{1}{2}\pm\frac{1}{\sqrt{5}}} \qquad f_{2,\pm} =\sqrt{\frac{1}{2}\pm\frac{2}{\sqrt{19}}}; 
\end{eqnarray}
finally, the Hamiltonian part of the dynamics is given by
\begin{equation}\label{37}
H = -71 \sqrt{2}\tau_1-\frac{\sqrt{2}}{3} \tau_2+23 \sqrt{2}\tau_3-12233\sqrt{\frac{2}{3}}\tau_8.
\end{equation}
The unraveling operator $A_{\psi}$ is hence directly defined by \reff{aucp},
while $B_{\psi,k}$ is obtained via the evaluation of the GTRO in \reff{transition}
and its diagonalization, see Eqs.~(\ref{eigen}) and (\ref{hb}).

Repeating the same calculations for the generator $\mathcal{G}^{CP}$ fixed by the full SA, i.e., \reff{eq:rcp}, we directly get
a diagonal form of the generator, with (positive) coefficients
\begin{equation}
c_1=c_2=c_3=c_4=c_5=4, \quad c_6 = \frac{8}{3}
\end{equation} 
and Lindblad operators, as well as the Hamiltonian, given by
\begin{eqnarray}
 L_1&= &\tau_3 \quad L_2=\tau_4 \quad L_3=\tau_5 \quad L_4=\tau_6 \nonumber\\ 
 L_5&=&\tau_7 \quad  L_6=\tau_8 \nonumber\\
 H&=& 23 \sqrt{2} \tau_3 -12233\sqrt{\frac{2}{3}}\tau_8.
\end{eqnarray}
Here, since the dynamics is CP one could apply the usual formulation of the (diffusive) unraveling, which is directly fixed by Eqs.~(\ref{aucp}) and (\ref{bucp}).

\begin{figure}[!t]
\includegraphics[width=0.5\columnwidth]{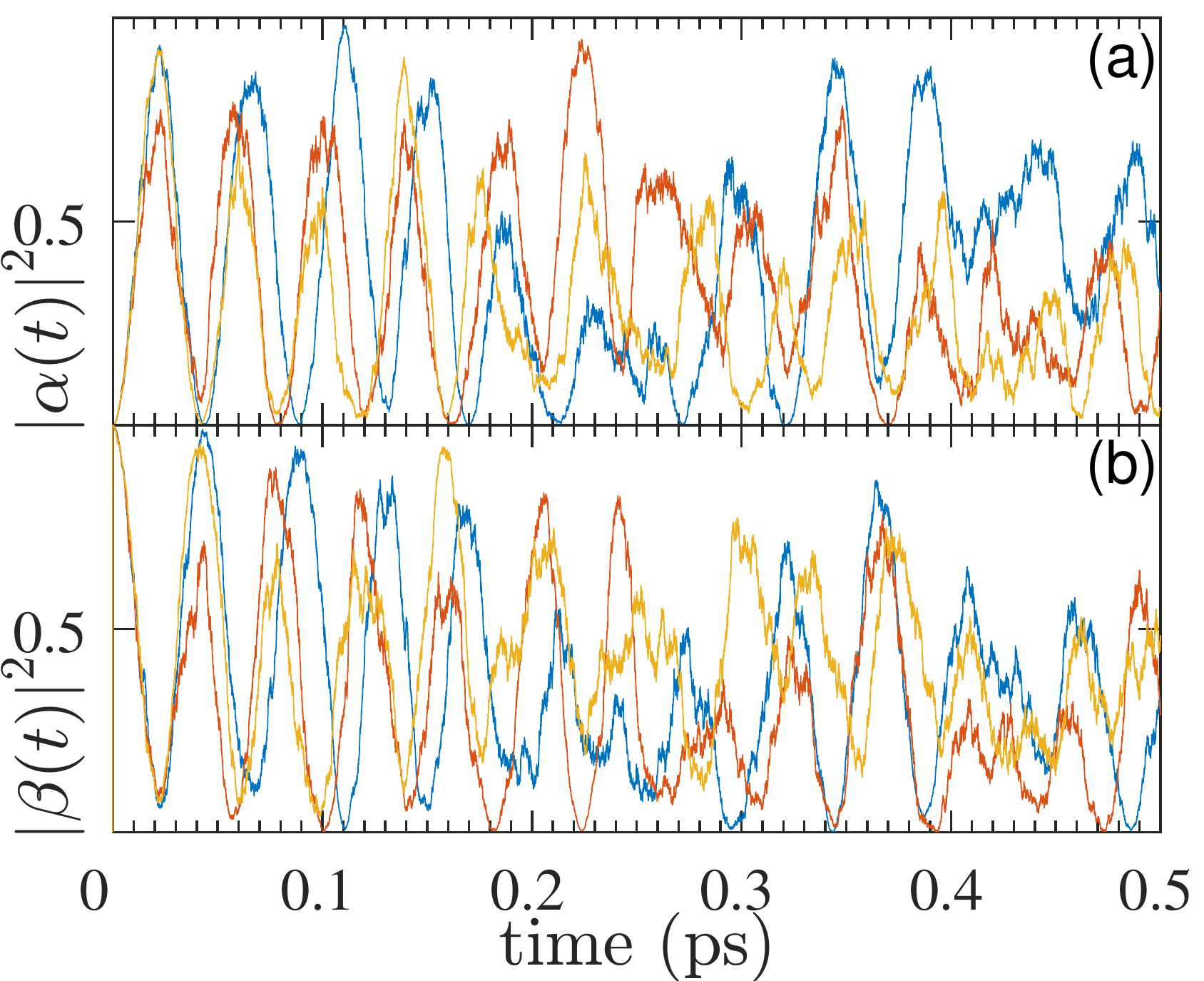}\hspace{0.cm}\includegraphics[width=0.5\columnwidth]{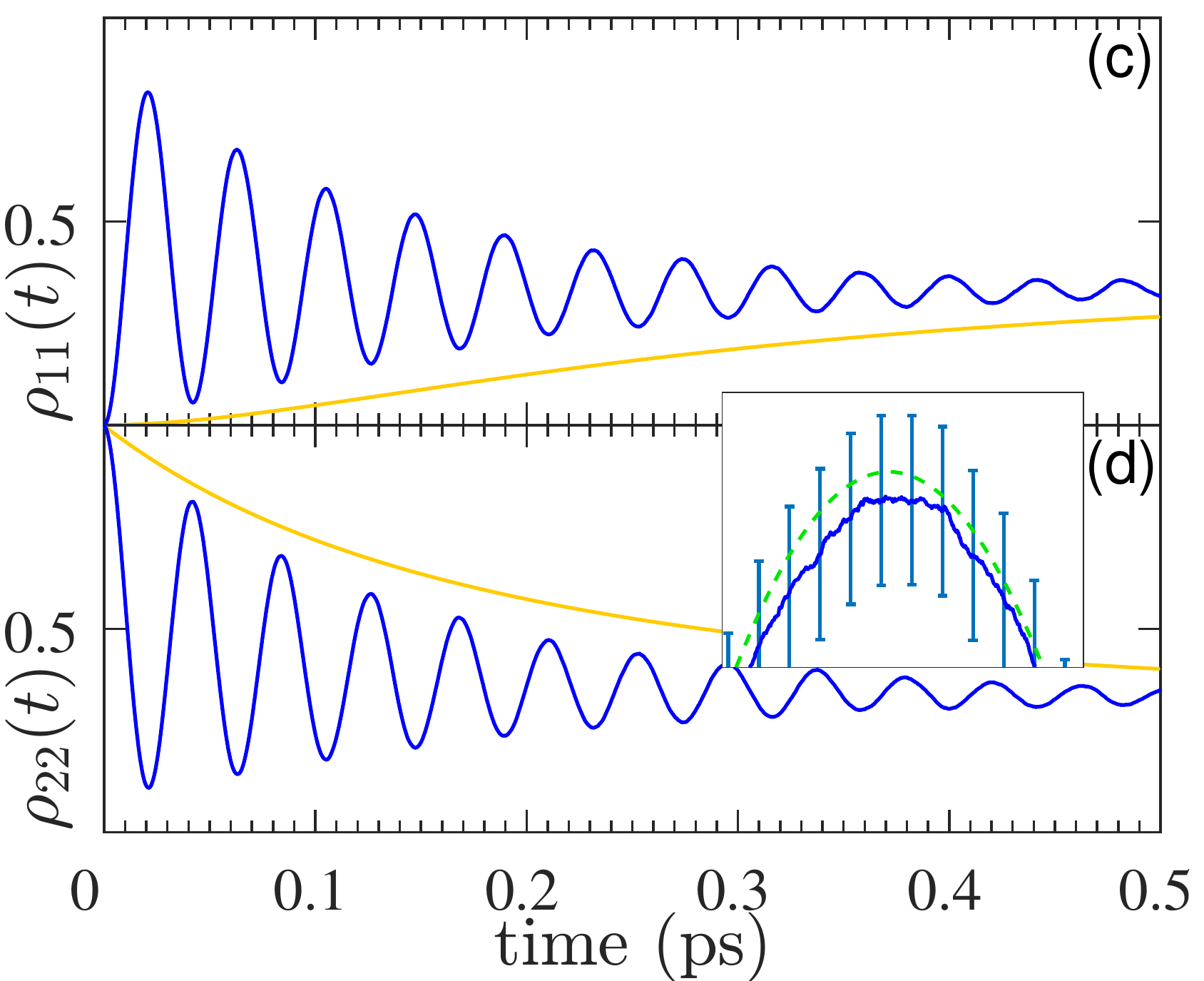}
\caption{Trajectories for the evolution of the population of site one \bf{(a)} and two \bf{(b)}; each trajectory corresponds to a different realization of the 
solution of the SDE in \reff{sde1}, with $A_{\psi_t} $ as in \reff{aucp} and $B_{\psi_t}$ as in \reff{hb} and
derived by diagonalizing the GTRO at each point of the computational time domain; the deterministic initial state
is $\ket{\psi(0)} = \ket{2}$, while the state at time $t$ is $\ket{\psi(t)}=\alpha(t)\ket{1}+\beta(t)\ket{2}+\gamma(t)\ket{3}$. 
Evolution of the population of site one \bf{(c)} and site two \bf{(d)} given by the ensemble average of $1000$ trajectories of our unraveling (blue/dark), and the solution of the Lindblad equation after full SA (yellow/light); 
in the inset, the ensemble average (blue/dark) and the solution of the P ME (green dotted) are shown to agree within the standard deviation of the mean (vertical bars) of the trajectories;
the initial state is set as $\rho(0)=\ket{2}\bra{2}$.}
\end{figure}

Now, our unravelling proceeds as the usual diffusive unravelling, with the 
addition that we have to diagonalize the GTRO in order to have the rate and noise operators
providing the trajectories. In particular, the algorithm giving each trajectory goes as follows. Firstly, $m=n-1$ Wiener processes with derivatives $d\xi_{k,t}$, with $k=0,\ldots,m$ and $n$ the dimension of the Hilbert space, are generated over a computational time domain $[0,\delta t,2\delta t,\ldots,T]$. Then, given the rates and the operators defining the ME (\ref{eq:eqje}) in the P case (see Eqs.~(\ref{35})-(\ref{37})), the expectation values $\ell_{\psi_0,j}$, the drift operators $A_{\psi_0}$ and the GTRO $W_{\psi_0}$ of \reff{aucp} and (\ref{transition}), respectively, are computed for a given initial state $\ket{\psi_0}$ at time $t_0$. Next, $W_{\psi_0}$ is diagonalized, the positivity of its eigenvalues is checked (a negative eigenvalue would stop the algorithm) and the noise operators $B_{\psi_0,k}$ are constructed according to \reff{hb}. Now, the state $\ket{\psi_{1}}$ after the first time step $\delta t$ is computed through the iterative formula $\ket{\psi_{1}}=\exp[(-iH+A_{\psi_0})\delta t+\sum_{i=1}^{m}B_{\psi_{t_0},i}d\xi_{i,0}]\ket{\psi_{0}}$ and then normalized. Finally, the state $\ket{\psi_0}$ is updated to $\ket{\psi_1}$ and the algorithm starts over with the evaluation of $\ell_{\psi_1,j}$, $A_{\psi_1}$ and $W_{\psi_1}$ at time $t_1=t_0+\delta t$. Once one trajectory is completed, another one is constructed starting with generating a new set of Wiener processes.

In Fig.1. \bf{a)} and \bf{b)} we report some trajectories for the evolution of the, respectively, first and second site populations,
which are obtained by means of the unraveling of the P dynamics after the partial SA, see Eqs.~\eqref{eq:eqje} and \eqref{eq:rp},
thus demonstrating the effectiveness of our approach on a physically relevant model.
Let us stress that the traditional unravelings for CP semigroups
could not be applied to these dynamics, since they
require a Lindblad equation and thus, in this context, a full SA.
Crucially, the latter would cancel any coupling between population and coherences, therefore potentially
disregarding some significant phenomena. This is explicitly shown in Fig.1. \bf{c)} and \bf{d)},
where we compare the evolution of the populations obtained by solving the Lindblad equation after the full SA and the 
populations obtained by averaging $1000$ trajectories of our unraveling. The
former completely neglects significant oscillations \cite{Jeske2015}, which are instead fully captured
by the unraveling of the P dynamics. 

As a final remark, we note that in the model at hand the P of the dynamics was guaranteed by itself. 
On the other hand, if we want to apply our unraveling to a more complex system,
starting from a generic ME as in \reff{diag} or in \reff{diagt}, which we do not know whether being P (divisible)
or not, we can still be sure that, as long as the algorithm works,
we are not dealing with ill-defined (i.e. non-positive) states.
Any detected phenomenon cannot be traced back to a non-physical description of the system's statistics. 
In fact, imagine that, on the contrary, the solution of a given ME maps the state of the system at a certain time $t$, $\rho(t)$,
into a non-positive state $\rho(t+\delta t)$.
If we now look at the unravelling, this means that  the GTRO will not be a positive operator,
for at least one of the stochastic states at time $t$ giving $\rho(t)$
on average. But then the algorithm will stop, due to the appearance of non-positive rates, see \reff{eigen},
witnessing the non-positivity of the map leading from time $t$ to $t+\delta t$.
This is fully analogous to what happens for, e.g., the non-Markovian quantum jumps
approach \cite{Piilo2008}, which can be safely applied to any ME, whose CP, or even P, may be not guaranteed. \\

\section{Conclusions}\label{sec:con} 
We have introduced a continuous unraveling for dynamics which are P, but not necessarily CP.
Our approach directly generalizes the QSD method: the rates and operators extracted from the ME have to be replaced by, respectively, the eigenvalues and eigenvectors of a proper rate operator.
We have taken into account the case of semigroup dynamical maps and, additionally, we have extended our result
to include a more general class of open-system evolutions,
so that our unravelling can be applied to every P-divisible dynamics.\\
By virtue of the unraveling of P dynamics, one can avoid to impose approximations which
could introduce significant errors in the system of interest, such as imposing the secular approximation
on top of the weak coupling approximation. This has been shown explicitly in a case study, by investigating
the population evolution in a dimer system.

Certainly, our approach can be improved in many regards.
A crucial point will be to simplify the task of diagonalizing the GTRO at each time step, e.g. by looking for possible connections between its spectral decompositions at subsequent times. 
Also, it will be of interest to study how and to what extent the range of applicability of our method
can be further extended, for example, combining it with other unraveling techniques \cite{Diosi1997,Gambetta2002,Piilo2008},
which apply to general non-Markovian dynamics.  
Finally, a central question, which is at the moment still open, is whether 
the unraveling we presented here can be formulated in terms of continuous measurements \cite{Wis96,Barchielli2009}.

\acknowledgments We thank Detlef D{\"urr} and G{\"u}nter Hinrichs for useful discussions and for their
valuable comments on an early version of this paper; we are also grateful to Lajos Di\'osi, Susana Huelga, Kimmo Luoma, Jyrki Piilo and Federico Carollo for helpful discussions. 
The work was financially supported by the John Templeton foundation (Grant No. 39530) and the QUCHIP Project (GA No. 641039).%

\end{document}